# N-gram Opcode Analysis for Android Malware Detection

**BooJoong Kang, Suleiman Y. Yerima, Sakir Sezer and Kieran McLaughlin**

*Centre for Secure Information Technologies (CSIT)*
*Queen's University Belfast, United Kingdom*
*Email: {b.kang, s.yerima, s.sezer, kieran.mclaughlin}@qub.ac.uk*

## ABSTRACT

Android malware has been on the rise in recent years due to the increasing popularity of Android and the proliferation of third party application markets. Emerging Android malware families are increasingly adopting sophisticated detection avoidance techniques and this calls for more effective approaches for Android malware detection. Hence, in this paper we present and evaluate an *n*-gram opcode features based approach that utilizes machine learning to identify and categorize Android malware. This approach enables automated feature discovery without relying on prior expert or domain knowledge for pre-determined features. Furthermore, by using a data segmentation technique for feature selection, our analysis is able to scale up to 10-gram opcodes. Our experiments on a dataset of 2520 samples showed an f-measure of 98% using the *n*-gram opcode based approach. We also provide empirical findings that illustrate factors that have probable impact on the overall *n*-gram opcodes performance trends.

## 1. INTRODUCTION

Android malware is an increasing problem due to its growing popularity and the ability of users to install applications from various application markets and third-party sources. The volume of new applications appearing frequently is too large for manual examination of each application for malicious behavior to be feasible. Hence, this process does not scale very easily to large numbers of applications. Previous studies have also shown that traditional signature-based approaches, which most antivirus scanners employ, fails to be effective at detecting new malware due to their increasing adoption of sophisticated detection avoidance techniques and the need for frequent update of signature databases.

Android malware detection is currently an active area of research. Consequently, there is a growing volume of work on automated detection incorporating machine learning techniques. Various methods have been proposed based on examining the dynamic application behavior (Zhao et al., 2011; Shabtai et al., 2012; Burguera et al. 2011), requested permissions (Liu & Liu, 2014; Sanz et al., 2012; Sharma & Dash, 2014; Chan & Song, 2014; Pehlivan et al., 2014; Rovelli & Vigfusson, 2014), API calls (Sharma & Dash, 2014; Yerima et al. 2015a; Yerima et al., 2015b; Dong-Jie et al., 2012; Chan & Song, 2014) etc. However these methods are often still largely reliant on expert analysis or domain knowledge to design or determine the discriminative features that are passed to the machine learning system used to make the final classification decision.

Some recent works have applied static opcode features to the problem of Android malware detection (Jerome et al, 2014; Kang et al., 2013; Canfora et al., 2015a; Canfora et al., 2015b; Varsha et al., 2016; Puerta et al., 2015; Canfora et al. 2015c). Out of these, only Jerome et al, (2014) and Canfora et al., (2015a) have investigated $n$-gram extracted from the disassembled application bytecode as a means for Android malware detection. The advantage of the use of an opcode based technique is the ability to automatically learn features from raw data directly rather than specifying them beforehand through expert analysis. For example, in previous works such as Yerima et al. (2015a) or Chan and Song (2014) where the discriminative features are based on API calls, expert analysis provides the selection of the 'most interesting' features (i.e. API calls methods, signatures, etc.). Unlike the opcode based techniques, this could limit the scope of the application of machine learning algorithms by excluding potentially useful learning information.

Hence, in this paper we investigate $n$-gram opcode analysis for Android malware detection using machine learning on real datasets. We study this approach and analyze its efficacy for both malware detection and also malware categorization (i.e. classification into known families). Unlike previous works that experimented with opcodes of up to 5-grams only, the work presented in this paper analyzed up to 10-grams whilst also considering both their frequencies and binary counts. We also some provide empirical findings that correlate with trends observed in the overall performance of the *n*-gram opcodes on the experimental dataset. The rest of the paper is organized as follows: Section 2 reviews related work; Section 3 presents the *n*-gram opcode analysis technique; the evaluation experiments and discussions are presented in Section 4 while Section 5 concludes the paper.

## 2. RELATED WORK

In this section we review related work on Android malware detection. The two main approaches commonly applied to malware detection are static analysis and dynamic analysis. Static analysis involves disassembling the application in order to extract features, while dynamic analysis involves running the application in an emulator or instrumented hardware in order to extract characteristic actions performed by the application. Static analysis has the advantage of being faster and may enable greater code coverage than dynamic analysis, while dynamic analysis may be less prone to code obfuscation. Some previous works such as Lindorfer, Neugschwandtner, and Platzer (2015); Titze, Stephanow, and Schütte (2013); Wei, Gomez, Neamtu, and Faloutsos (2012) have combined the two approaches.

Learning based approaches using hand-designed (pre-defined) features have been applied extensively to both dynamic (Zhao et al., 2011; Shabtai et al., 2012; Burguera et al., 2011; Su et al., 2012; Dini et al., 2012; Afonso et al., 2015) and static malware detection (Sharma & Dash, 2014; Yerima et al. 2015a; Yerima et al., 2015b; Dong-Jie et al., 2012; Chan & Song, 2014). For example, Yerima et al. (2015a) studied a static analysis approach to Android malware detection based on 179 features derived from API calls, intents, permissions and commands that were combined with ensemble learning. Their approach was evaluated on a dataset of 2925 malware samples and 3938 benign samples. A variety of similar approaches to static malware detection have used similarly derived features, but with different classifiers such as support vector machine (SVM) (Arp et al., 2014), Naïve Bayes (Yerima et al., 2014), and *k*-nearest neighbor (KNN) (Sharma and Dash, 2014).

Chan and Song (2014) extracted some pre-defined API calls categorized into privacy related, network related, SMS related, Wi-Fi related and components related API calls. They combined these features with Android permissions and trained 7 classifiers: Naïve Bayes, SVM, radial basis function (RBF), multi-layer perceptron (MLP), Liblinear, decision tree and random forest. They found that better classification accuracy was obtained by the random forest classifier when API calls were used with permissions compared to the use of permissions alone. Their experiments were performed on a dataset of 796 samples consisting of 621 benign and 175 malware samples.

Machine learning based Android malware detection approaches have also been proposed that use static features derived exclusively from permissions. Liu et al. (2014) proposed a 2-layer permissioned based Android malware detection scheme. In their system, each stage utilizes a J48 decision tree algorithm to identify malicious applications. The first layer uses requested permissions and requested permission pairs, while the second layer employs 'used permission pairs' extracted from Dalvik executable files. A simple three-step algorithm is used to obtain the overall classification decision. Sanz et al. (2012) evaluated simple logistic, Naïve Bayes, Bayes network, SVM, IB$k$ (KNN), J48, random tree and random forest classifiers using permissions as features. They experimented on 249 malware samples and 357 benign samples and obtained the best results with the random forest classifier (AUC=0.92).

Pehlivan et al. (2014) presented experimental results on 2338 benign and 1446 malware samples using permissions. They investigated 4 different feature selection methods with Naïve Bayes, classification and regression tree (CART), J48, SVM and random forest classifiers. Their results also showed that the random forest classifier exhibits the best classification accuracy (94.9%). Rovelli and Vigfusson (2014) presented PMBS, a permission-based malware detection system. PMBS extracts requested permissions as behavior markers and builds machine learning classifiers to automatically identify potentially harmful behavior. They considered C4.5, lazy instance based learner, repeated incremental pruning to produce error reduction (RIPPER) and Naïve Bayes classifiers and experimented with 1500 benign and 1450 malware samples. They also applied the boosting meta-learning technique on the C4.5, RIPPER and Naïve Bayes classifiers. Their results showed that the 'boosted' C4.5 performed best with 95.22% accuracy. APK Auditor (Talha, Alper & Aydin, 2015) uses permissions with a statistical scoring approach to detect malicious Android applications.

In contrast with the aforementioned approaches that rely on high-level pre-defined features, such as permissions or API calls, $n$-grams based malware detection uses sequences of low-level opcodes as features. The $n$-gram features can be used to train a classifier to distinguish between malware and benign applications (Jerome et al., 2014), or to classify malware into different families (Kang et al., 2013). Perhaps surprisingly, even 1-gram based features, which are simply a histogram of the number of times each opcode is used, can be useful in distinguishing malware from benign applications (Canfora et al., 2015b). The length of the $n$-gram used (Jerome et al., 2014) and number of *n*-grams used in classification (Canfora et al., 2015b) can both have an effect on the accuracy of the classifier.

Jerome et al. (2014) utilized opcode sequences with machine learning and experimented with 2, 3, 4 and 5-gram binary features. Feature ranking and selection were done based on computing the information gain of each *n*-gram of opcode sequences found in the training sets. The authors assessed their approach on the Android malware genome project (AMGP) samples, and benign samples obtained from Google Play. They were able to achieve the best classification performance with the 5-gram features (average global F-measures close to 0.9771) by utilizing a linear SVM classifier. The paper only investigated malware classification into benign or suspicious and did not consider malware family classification. Moreover, the approach is based on binary occurrences of *n*-gram opcodes within each application and did not consider any frequency information associated with the *n*-gram opcodes found in each application.

Unlike Jerome et al. (2014), Canfora et al. (2015a) presented results based on malware family classification to test their approach using (Gaussian) SVM and random forest classifiers. They presented results on evaluation with a set of trusted applications and also 10 Android malware families: FakeInstaller, Plankton, DroidKungFu, GinMaster, BaseBridge, ADRD, Kmin, Gemini, DroidDream, and Opfake. Furthermore, their experiments covered ranges of *n* from 1 to 5, i.e. up to a maximum of 5-gram opcodes and up to 2000 *n*-gram features per scheme after applying a feature selection algorithm. The best result reported in the paper is 96.88 % accuracy using the SVM classifier with 2-gram features numbering 1000. Nevertheless, the authors noted that higher gram features likely performed worse than 2-grams because of the number of features used were limited to 2000 only.

Note that although Kang et al. (2013), Canfora et al. (2015a), Puerta et al. (2015), Varsha et al. (2016) and Canfora et al. (2015c) utilized static opcode features, they were not based on *n*-grams. *N*-grams have recently been applied to system call sequences obtained from Android applications by Mas'ud et al. (2016). However, static opcode sequences have much less computational overhead compared to system call sequences since the latter is based on dynamic analysis. Moreover a static feature extraction approach covers much more code than a dynamic approach where code coverage is an ongoing problem. Thus, some malware functionality may not be triggered leading to incomplete collection of relevant system call sequences.

In this paper, a static *n*-gram opcode based approach is also investigated. However, unlike previous works we develop an approach that enables the use of longer *n*-grams thus analyzing up to 10-grams whereas the previously reported works utilized up to 5-grams (Jerome et al., 2014 and Canfora et al., 2015a). Furthermore, unlike the work by Jerome et al. (2014) which was only based on binary information, we investigate both frequency and binary information thus allowing for greater information coverage. Finally, we also present some empirical findings that correlate with the overall observed performance trends of the *n*-gram opcodes on the experimental dataset used.

## 3.    N-GRAM OPCODE ANALYSIS

In this section, we explain how to extract *n*-gram opcodes from Android applications and how to select *n*-gram opcodes that will enable optimal malware detection and categorization. The following two subsections describe the process of the *n*-gram opcode extraction and the feature selection and also include statistical results from our dataset.

## 3.1.    N-gram opcode extraction

The *n*-gram opcode extraction consists of disassembling applications and extracting *n*-grams from opcode sequences. An Android application can be delivered as a compressed file, an Android application package (apk) file, containing a manifest file, resource files and Dalvik executable (dex) files. The dex files contain the application bytecode and can be disassembled using baksmali (http://baksmali.com). As a result of disassembling, baksmali generates a set of smali files for the dex file, where each smali file represents a single class that contains all the methods of the class. Each method contains human-readable Dalvik bytecode (hereafter instructions) and each instruction consists of a single opcode and multiple operands.

We discard the operands and only extract *n*-grams from opcode sequences of the methods. The resulting output of the *n*-gram opcode extraction is a vector of unique *n*-gram opcodes from all the classes of the application. The vector contains the frequency of each unique *n*-gram opcode. The overview of the *n*-gram opcode extraction is shown in Figure 1 and there are five main groups of opcodes obtainable from the Dalvik bytecode (Bartel, Klien, Le Traon, & Monperrus, 2012). These include:

- Move instructions (0x01 to 0x1C)
- Branch instructions (0x27 to 0x3D)
- Getter and setter instructions (0x44 to 0x6D)
- Method invoke instructions (0x6E to 0x78)
- Logic and arithmetic instructions (0x7B to 0xE2)

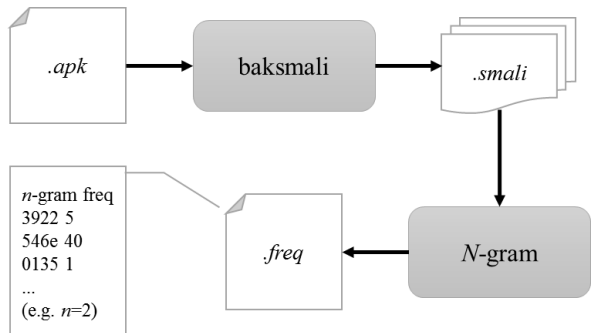

*Figure 1. N-gram opcode extraction process.*

Any *n*-gram based method faces the prospect of exponential increase in the number of unique *n*-grams as *n* is increased. Hence, it was expected that a similar trend would be observed with *n*-gram opcodes as well. Figure 2 and Table 1 show the number of unique *n*-gram opcodes for different *n* values from our dataset, which consists of 1260 samples of malware and 1260 benign samples. In the malware detection (MD) study, we processed the *n*-gram opcode extraction on all the 2520 samples and counted the number of unique *n*-opcodes for different *n*, with *n* ranging from 1 to 10. In the malware categorization (MC) study, we conducted the same process on 1260 malware samples.

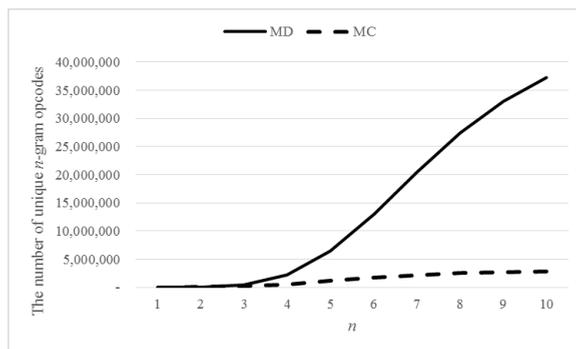

*Figure 2. The number of unique n-gram opcodes vs. n.*

It can be observed that there is no exponential increase in the number of unique *n*-gram opcodes. Instead, it increases linearly at first and then gradually becomes stable. A probable reason for this is that some methods may contain less instructions than *n* therefore those *n*-gram opcodes will not appear in such methods. Another reason is that a bigger *n* is likely to generate a smaller number of *n*-gram opcodes than a smaller *n* generates from the same method. For example, a method with 7 instructions has 6 pieces of 2-gram opcodes, 5 pieces of 3-gram opcodes, 4 pieces of 4-gram opcodes and so on. Therefore, the maximum number of unique *n*-gram opcodes for a method is in inverse proportion to *n*. In the MC study, the number of unique *n*-gram opcodes also becomes stable despite utilizing only malware samples for the study. Even though there is no exponential increase, nonetheless an excessive number of unique *n*-gram opcodes that could cause a huge overhead in further processes results. The number of the unique 10-gram opcodes in MD and MC were observed to be 37,186,183 (about 37M) and 2,823,729 (about 2.8M), respectively.

*Table 1. The number of unique n-gram opcodes for different values of n.*

| *n* | *MD* | *MC* |
| --- | --- | --- |
| 1 | 214 | 210 |
| 2 | 22,371 | 14,550 |
| 3 | 399,598 | 154,483 |
| 4 | 2,201,377 | 557,526 |
| 5 | 6,458,246 | 1,145,025 |
| 6 | 12,969,857 | 1,724,771 |
| 7 | 20,404,473 | 2,177,621 |
| 8 | 27,366,890 | 2,491,721 |
| 9 | 33,024,116 | 2,695,226 |
| 10 | 37,186,183 | 2,823,729 |

## 3.2. Feature selection

Since the number of unique *n*-gram opcodes is excessive, it is difficult to run machine learning algorithms on the original data. A solution is feature selection i.e. a process of identifying the best features and is a widely used approach to filter out less important features. In the feature selection stage, we measure the information gain of each feature and subsequently filter out the less important features that have low information gain.

Entropy is a measure of the uncertainty of a random variable. The entropy of a variable *X* is defined in (1) below and the entropy of *X* after observing values of a variable *Y* is defined in (2), where $P(x_i)$ is the prior probabilities for all values of *X* and $P(x_i|y_i)$ is the posterior probabilities of *X* given the values of *Y*.

(1) $$H(X) = -\sum P(x_i)\log_2(P(x_i))$$

(2) $$H(X/Y) = -\sum P(y_j) \sum P(x_i/y_j)\log_2(P(x_i/y_j))$$

The amount by which the entropy of *X* decreases reflects additional information about *X* provided by *Y* and is called information gain (Quinlan, 1993), given by (3).

(3)                   $IG(X/Y) = H(X) - H(X/Y)$

According to this measure, a feature *Y* is regarded as a better indicator than a feature *Z* for a class *X*, if $IG(X|Y) > IG(X|Z)$. We rank the features by the information gain and select the high ranked features. In order to compute the *IG*, we used an implementation of the information gain in WEKA (Hall et al., 2009). However, the program could not handle the large data input from the *n*-gram opcode feature files and frequently encountered an out of memory error (on a Linux PC with 32GB RAM). So in order to overcome this problem, we segmented the data into several smaller chunks (in multiple .arff files) and computed the information gain on smaller data. This worked because the information gain algorithm computes a score for each feature independently. Hence, we processed the information gain on each smaller set of features and merged the results together at the end as illustrated in Figure 3. This 'data segmentation' approach to the feature selection allowed us to overcome the memory limitation problem and thus experiment on larger *n*-gram opcode features.

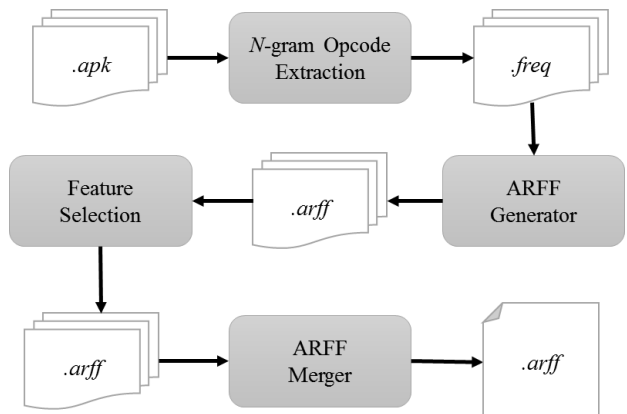

*Figure 3. Overview of our analysis process illustrating feature segmentation applied to the feature selection stage.*

Figure 4 and Table 2 show the number of selected *n*-gram opcodes with the information gain greater than 0.1 for different *n*. As the table illustrates, the number of selected *n*-gram opcodes increased as *n* becomes larger. This means that we gain more information by increasing *n*. However, the increase in the number of selected *n*-gram opcodes reaches a saturation point as *n* increases. Because of this observation, we expect that impact of the increase in *n* on classification accuracy to peak at this saturation point where further increase in *n* will have little effect. Another interesting observation is that the number of selected *n*-gram opcodes for frequency is slightly higher than the number of selected *n*-gram opcodes for binary. Note that frequency and binary refer to the number of counts of the *n*-gram opcode. For the former it is the overall count, while for the latter it is 1 or 0 denoting presence and absence respectively.

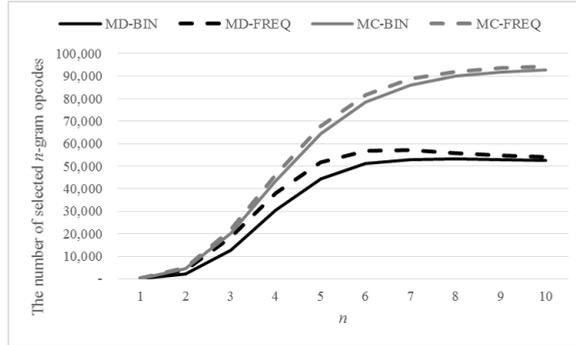

*Figure 4. The number of selected n-gram opcodes vs. n.*

*Table 2. The number of selected n-gram opcodes.*

| n | MD | | MC | |
|---|---|---|---|---|
| | Binary | Frequency | Binary | Frequency |
| 1 | 29 | 191 | 164 | 199 |
| 2 | 2,121 | 4,155 | 4,567 | 4,973 |
| 3 | 12,802 | 18,411 | 20,376 | 21,851 |
| 4 | 30,293 | 37,992 | 43,538 | 46,253 |
| 5 | 44,532 | 51,845 | 64,450 | 67,837 |
| 6 | 51,213 | 56,850 | 78,453 | 81,736 |
| 7 | 53,079 | 57,086 | 86,146 | 88,839 |
| 8 | 53,139 | 55,856 | 90,024 | 92,076 |
| 9 | 52,857 | 54,837 | 91,966 | 93,600 |
| 10 | 52,588 | 54,080 | 92,878 | 94,239 |

## 4. EVALUATION

In this section, we evaluate the performance of *n*-gram opcodes for malware detection and categorization with different *n*. Evaluation and detailed analyses for malware detection and categorization are presented in the following subsections. In each subsection, we also compare the performance of two different data types: binary and frequency.

Our dataset consists of malware from the Android malware genome project (Zhou & Jiang, 2012) and has a total of 2520 applications, of which 1260 are benign and 1260 are malware from 49 different malware families. Labels are provided for the malware family of each sample. The benign samples were collected from the Google Play store and have been checked using VirusTotal (http://www.virustotal.com) to ascertain that they were highly probable to be malware free. We use four different machine learning algorithms: Naïve Bayes (NB), support vector machine (SVM), partial decision tree (PART) and random forest (RF) and utilize WEKA as the framework. The following experimental results are reported using the weighted average f-measure, which is based on the precision and recall, over 10-fold cross validation.

## 4.1.    Malware detection

In the malware detection study, samples are classified into one of two classes: benign or malware. We evaluated the performance of malware detection with two different data types of $n$-gram opcodes: binary and frequency.

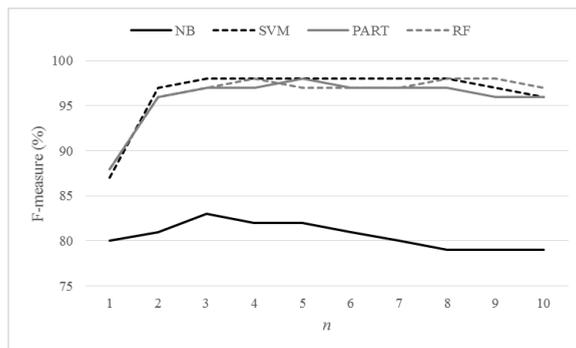

*Figure 5. Malware detection results with binary n-gram opcodes.*

Figure 5 shows the results of the binary $n$-gram opcodes. With the exception of NB, the performances of the other three algorithms were similar although SVM shows the best performance in most cases. The f-measure increases as $n$ is increased but no more increase is observed when $n$ is greater than 3. The f-measure even tends to decrease when $n$ is greater than 7. This trend is similar with the change in the number of selected $n$-gram opcodes in Figure 4. NB shows the worst performance but shows the same trend. The best f-measure is 98% and SVM performed with the best f-measure when $n$ is 3.

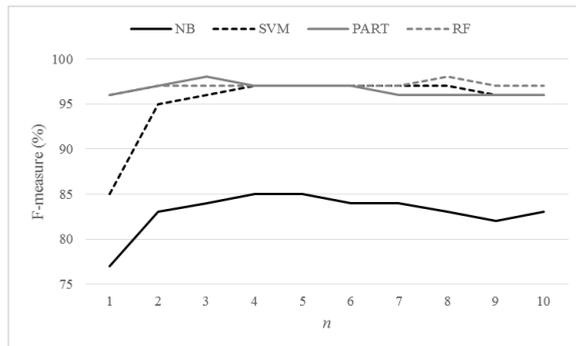

*Figure 6. Malware detection results with frequency n-gram opcodes.*

As shown in Figure 6, the frequency *n*-gram opcodes show similar results with the binary *n*-gram opcodes. One interesting observation is that the frequency *n*-gram opcodes show a good performance when *n* is 1 (i.e. no *n*-gram is applied) compared to the binary *n*-gram opcodes.

*Table 3. Top ten binary n-gram opcodes for malware detection.*

| Rank | Our Findings | | | Jerome (2014) |
|------|------|------|------|------|
| | *3* | *4* | *5* | *5* |
| 1 | **08076e** | 6e0c086e# | 6e0c086e0c# | 136e6e6e0c |
| 2 | 08546e# | 08546e0c# | 2038071f6e# | 1c6e6e0c6e* |
| 3 | 0c086e | 0c086e0c# | 0854380854# | 1d546e0a39 |
| 4 | 220870# | 08546e0a# | 38071f6e0a# | 210135461a |
| 5 | 12086e# | 08540854# | 0854085408# | 0c1a6e0a33 |
| 6 | 390f6e# | 38071f6e# | 0c08546e0c# | 12123c0e22* |
| 7 | 085412# | 390f6e0a# | 0c086e0c6e# | 7154626e28 |
| 8 | 390f54 | 08543808# | 08546e0854# | 6e2820381f |
| 9 | 085208# | **08076e0c** | 20381f2822 | 0d076e289c* |
| 10 | 085438# | 3808546e# | 08546e0c08# | 16313d740e* |

Table 3 shows the top ten binary *n*-gram opcodes for malware detection (see the translation at Appendix). We also present the top ten 5-gram opcodes of Jerome et al. (2014) (which were the only ones provided) to compare with our findings. Even though no same 5-gram opcodes commonly appear between the two top ten 5-gram opcodes (in the two rightmost columns), we found six of the top ten 5-gram opcodes from Jerome et al. (2014) , highlighted in the last column, in our selected 5-gram opcodes as the 11770th, 12597th, 13241th, 13260th, 15927th and 25662th ranked features, respectively. This could be attributed to the fact that our benign samples differed even though we utilized the same malware samples as Jerome et al. (2014). An interesting observation is that three of the overlapping *n*-gram opcodes ('12123c0e22', '0d076e289c' and '16313d740e') are also only found in malware samples just as observed by Jerome et al. (2014). Another observation is that our top hundred 5-gram opcodes are mostly found in benign samples. As we highlighted *n*-gram opcodes only found in benign samples with '#', it can be seen from the table that there were more of the top ten *n*-gram opcodes only found in benign samples. This would indicate that the unique *n*-gram opcodes from benign samples were a strong contributing factor for malware detection.

Another interesting observation is that three 4-gram opcodes ('08546e0c', '08546e0a' and '3808546e') were extensions of the second ranked 3-gram opcode '08546e'. We refer to these extensions as extended *n*-gram opcodes, which includes at least one (*n-1*)-gram opcode as a prefix or suffix. For example, '08546e0c' includes '08546e' as a prefix and '3808546e' includes '08546e' as a suffix. If an *n*-gram opcode does not include any (*n*-1)-gram opcode as a prefix or suffix, we refer to this *n*-gram opcode as a new *n*-gram opcode. A (*n*-1)-gram opcode usually produces multiple extended *n*-gram opcodes which provide more precise information of opcode sequences. In contrast, new *n*-gram opcodes provide the information of new opcode sequences. In order to investigate the impact of new and extended *n*-gram opcodes on the classification accuracy, additional analysis has been conducted as follows.

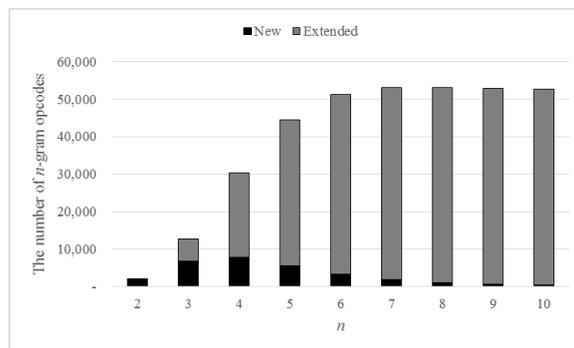

*Figure 7. The number of new and extended n-gram opcodes for binary.*

Figure 7 shows the number of new and extended *n*-gram opcodes for the binary data type. The number of new *n*-gram opcodes increased until *n* reaches 4 and them decreased with higher values of *n*. It means that we extracted less information of new opcode sequences at high values of *n*. This observation might be one of reasons why the classification accuracy has been observed to saturate at lower values of $n$, i.e. from 2 to 4, even though the number of selected $n$-gram opcodes was still on the increase. Note that the number of selected $n$-gram opcodes saturated when $n$ was 6 or 7, but the classification accuracy saturated when $n$ was 3 or 4 (see Figure 4 and 5). Figure 8 supports this argument as well, by showing that the proportion of new $n$-gram opcodes became less than the proportion of extended $n$-gram opcodes.

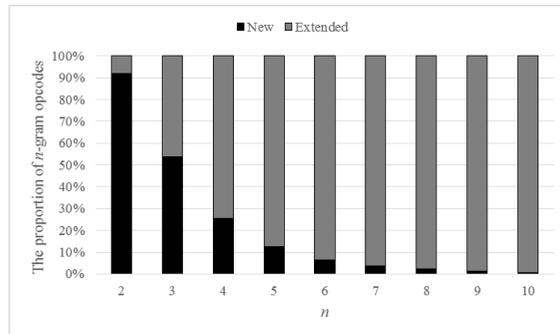

*Figure 8. The proportion of new and extension n-gram opcodes for binary.*

As explained before, extended $n$-gram opcodes include one $(n-1)$-gram opcode as a prefix or suffix. However, there is a special case that an extended $n$-gram opcode includes two $(n-1)$-gram opcodes both as a prefix and a suffix. For example, '3808546e' includes '380854' as a prefix and '08546e' as a suffix. We call these extended $n$-gram opcodes as overlaps of two $(n-1)$-gram opcodes. According to Figure 9 and 10, the number and the proportion of the overlaps were larger than those of prefix and suffix when $n$ was higher than 4. This analysis shows that most of selected $n$-gram opcodes were extended $n$-gram opcodes and most of extended $n$-gram opcodes were overlaps when the classification accuracy has been saturated.

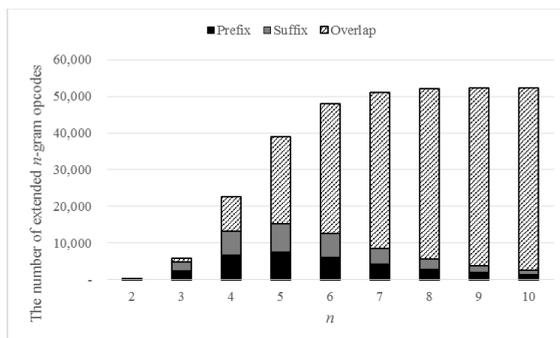

*Figure 9. The number of each type of extension: prefix, suffix, or overlaps (for binary).*

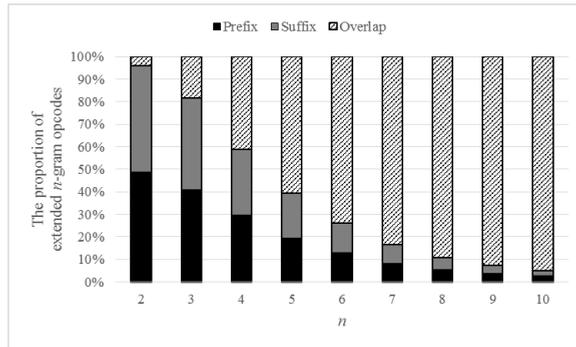

*Figure 10. The proportion of each type of extension: prefix, suffix or overlaps (for binary).*

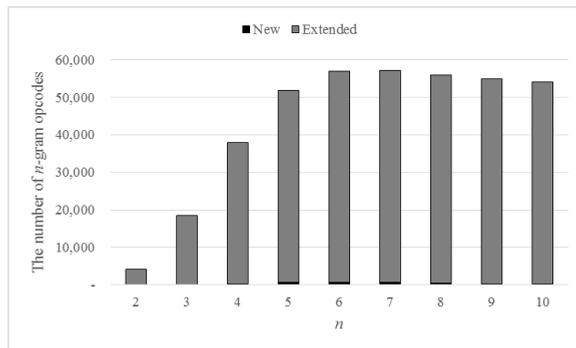

*Figure 11. The number of new and extension n-grams (for frequency).*

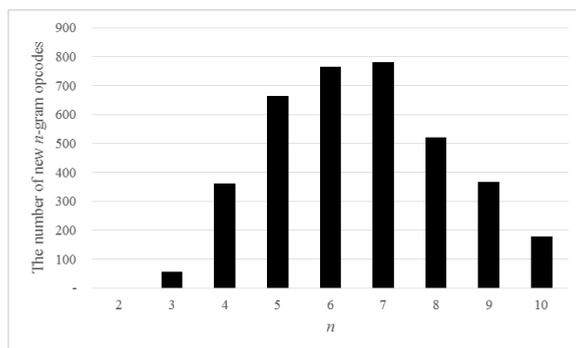

*Figure 12. The number of new n-grams (for frequency features).*

As shown Figure 11 and 12, the number and proportion of new $n$-gram opcodes were very small compared to those of extended $n$-gram opcodes. Similar to the case of the binary data type, most of extended $n$-gram opcodes were also overlaps (see Figure 13). Note that the classification accuracy of the frequency data type becomes saturated at an earlier point compared to the detection accuracy of the binary type (see Figures 5 and 6).

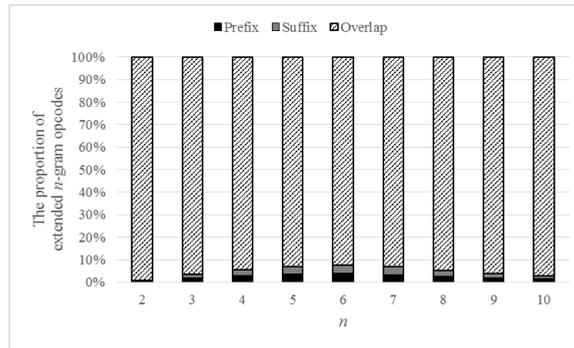

*Figure 13. The number of each type of extension: prefix, suffix or overlaps (for frequency).*

In terms of classification accuracy, SVM showed the best performance in the most cases. However, the accuracy is not the only factor when choosing a machine learning algorithm. Another factor is the speed performance and Table 4 shows the time overhead of the machine learning algorithms from the malware detection with the binary $n$-gram opcodes. RF showed the best performance in terms of both training and prediction speeds. Even though PART has the highest training overhead it is still considered a suitable classifier because once trained, classification is fast as well.

*Table 4. Time overhead of the machine learning algorithms.*

| $n$ | Machine Learning Algorithms | | | |
|---|---|---|---|---|
| | NB | SVM | PART | RF |
| 1 | 0.02/0.01 | 1/0 | 0.14/0 | 0.14/0 |
| 2 | 0.31/0.22 | 1.53/0.01 | 3.26/0 | 0.27/0 |
| 3 | 2.18/1.5 | 5.65/0.03 | 25.82/0 | 0.34/0.01 |
| 4 | 6.41/3.36 | 12.8/0.1 | 62.38/0.01 | 0.51/0.01 |
| 5 | 11.34/5.02 | 19.17/0.25 | 106.13/0.02 | 0.74/0.02 |
| 6 | 14.66/5.96 | 22.32/0.32 | 191.06/0.02 | 0.96/0.02 |
| 7 | 14.23/6.24 | 24.45/0.38 | 176.92/0.02 | 1.15/0.02 |
| 8 | 13.42/5.97 | 21.17/0.38 | 169.25/0.02 | 1.88/0.02 |
| 9 | 13.05/5.99 | 21.29/0.39 | 157.95/0.01 | 4.17/0.02 |
| 10 | 13.36/5.95 | 27.26/0.36 | 159.1/0.01 | 6.07/0.02 |

## 4.2. Malware categorization

In the malware categorization study, samples are classified into one of existing malware families. We also evaluated the performance of malware categorization with two different data types of $n$-gram opcodes. Figure 14 shows the results of binary $n$-gram opcodes which depicts a similar trend to the malware detection results. SVM shows the best f-measure of 98%, with its f-measure becoming steady when $n$ is 4.

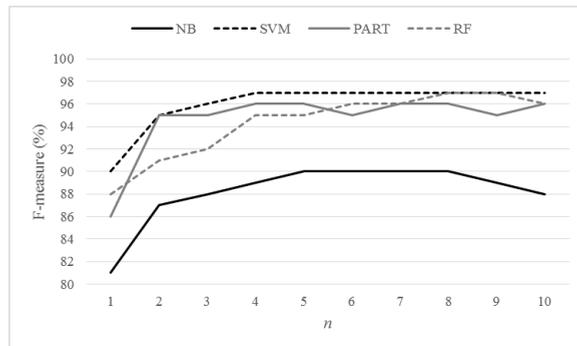

*Figure 14. Malware categorization results with binary n-gram opcodes.*

Figure 15 shows the results of frequency $n$-gram opcodes for malware categorization. Again, SVM shows the best f-measure of 98%, when $n$ is 6 but SVM and PART show a high f-measure of 95%, when $n$ reaches only 2. The frequency 1-gram opcodes do not show a better performance compared to the binary 1-gram opcodes, as it does in malware detection. In contrast to malware detection, malware categorization is a multi-class classification, which is considered to be a more difficult problem compared to the binary classification.

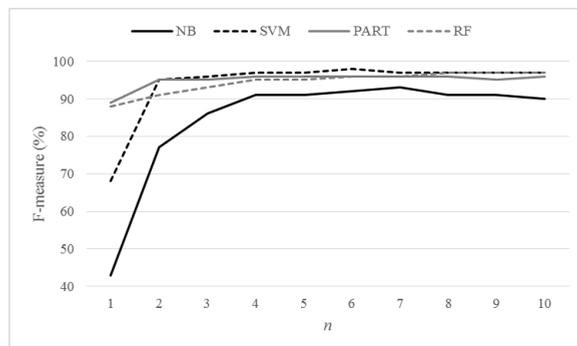

*Figure 15. Malware categorization results with frequency n-gram opcodes.*

Table 5 shows the top ten binary *n*-gram opcodes for malware categorization. Since we have no data from other research for comparison, we only analyze our data for malware categorization. We observed that there is no *n*-gram opcode that was found only (exclusively) in a single family within the top ten *n*-gram opcodes.

*Table 5. Top ten frequency n-gram opcode for malware categorization.*

| Rank | *n* | | |
|------|-----|-----|-----|
| | *3* | *4* | *5* |
| 1 | 36121a | 700c390e | 22621a701a |
| 2 | 3c0e22 | 621a701a | 621a701a71 |
| 3 | 1a1607 | 123c0e22 | 123c0e2270 |
| 4 | 616174 | 12123c0e | 3c0e22706e |
| 5 | 3b7428 | 313b7428 | 12123c0e22 |
| 6 | 313b74 | 3c0e2270 | 0c6e0c236e |
| 7 | 123c0e | 1a706e15 | 6e0a386e71 |
| 8 | 3d740e | 3922701c | 0c22702271 |
| 9 | 0a8104 | 3d740e0d | 0c1a221a6e |
| 10 | 289c08 | 8104085a | 0c1a706e15 |

In our evaluation, it is apparent that the binary *n*-gram opcodes are more accurate than the frequency *n*-gram opcodes. Another advantage of the binary *n*-gram opcode is that we can reduce the storage overhead as mentioned in (Jerome et al., 2014). However, the frequency *n*-gram opcodes show very good accuracy when *n* is small. This means that the frequency *n*-gram opcodes with small *n* can be chosen for light-weight use case scenarios.

## 5.  CONCLUSION

In this paper, we investigated and analyzed *n*-gram opcode based static analysis approach to Android malware detection. This approach eliminates the need for prior expert or domain knowledge based features. Unlike most previous works that utilize pre-defined features like API calls, permissions, intents and other application properties, our method allows for automatic extraction and learning of features from given datasets. Furthermore, we achieved analysis with longer *n*-grams than the state-of-the art by utilizing up to 10-gram opcodes in our experiments compared to the currently reported maximum of 5 in the literature. This was possible using a data segmentation technique during pre-processing in order to enable feature selection on as large a dataset as possible. Our results showed that by using frequency $n$-gram opcodes with low $n$, good classification accuracy can be achieved. Nevertheless, a maximum f-measure of 98% in both malware detection and categorization were obtained with $n=3$ and $n=4$, respectively. We also showed that the presence of overlapping and extended $n$-gram opcodes were in correlation with the overall detection accuracy results. Nevertheless, we intend to investigate this aspect further. For future work, experimenting on larger datasets would be considered. We would also like to investigate the performance of mixed feature sets comprising different lengths of $n$-grams, especially those that do not overlap across different $n$ values.

## APPENDIX

| 1.     08 07 6e | 6.    39 0f 6e |
|---|---|
| move-object/from16 | if-nez |
| move-object | return |
| invoke-virtual | invoke-virtual |
| 2.     08 65 6e | 7.    08 54 12 |
| move-object/from16 | move-object/from16 |
| sget-char | iget-object |
| invoke-virtual | const/4 |
| 3.     0c 08 6e | 8.    39 0f 54 |
| move-result-object | if-nez |
| move-object/from16 | return |
| invoke-virtual | iget-object |
| 4.     22 08 70 | 9.  08 52 08 |
| new-instance | move-object/from16 |
| move-object/from16 | iget |
| invoke-direct | move-object/from16 |
| 5.     12 08 6e | 10.    08 54 38 |
| const/4 | move-object/from16 |
| move-object/from16 | iget-object |
| invoke-virtual | if-eqz |

*Table 6. Top 10 3-gram opcodes from malware detection.*

| 1.     6e 0c 08 6e | 6.     38 07 1f 6e |
|---|---|
| invoke-virtual | if-eqz |
| move-result-object | move-object |
| move-object/from16 | check-cast |
| invoke-virtual | invoke-virtual |
| 2.     08 54 6e 0c | 7.     39 0f 6e 0a |
| move-object/from16 | if-nez |
| iget-object | return |
| invoke-virtual | invoke-virtual |
| move-result-object | move-result |
| 3.     0c 08 6e 0c | 8.     08 54 38 08 |
| move-result-object | move-object/from16 |
| move-object/from16 | iget-object |
| invoke-virtual | if-eqz |
| move-result-object | move-object/from16 |

| 4.  08 54 6e 0a | 9.  08 07 6e 0c |
|---|---|
| move-object/from16 | move-object/from16 |
| iget-object | move-object |
| invoke-virtual | invoke-virtual |
| move-result | move-result-object |
| 5.  08 54 08 54 | 10.  38 08 54 6e |
| move-object/from16 | if-eqz |
| iget-object | move-object/from16 |
| move-object/from16 | iget-object |
| iget-object | invoke-virtual |

*Table 7. Top 10 4-gram opcodes from malware detection.*

| 1.  6e 0c 08 6e 0c | 6.  0c 08 54 6e 0c |
|---|---|
| invoke-virtual | move-result-object |
| move-result-object | move-object/from16 |
| move-object/from16 | iget-object |
| invoke-virtual | invoke-virtual |
| move-result-object | move-result-object |
| 2.  20 38 07 1f 6e | 7.  0c 08 6e 0c 6e |
| instance-of | move-result-object |
| if-eqz | move-object/from16 |
| move-object | invoke-virtual |
| check-cast | move-result-object |
| invoke-virtual | invoke-virtual |
| 3.  08 54 38 08 54 | 8.  08 54 6e 08 54 |
| move-object/from16 | move-object/from16 |
| iget-object | iget-object |
| if-eqz | invoke-virtual |
| move-object/from16 | move-object/from16 |
| iget-object | iget-object |
| 4.  38 07 1f 6e 0a | 9.  20 38 1f 28 22 |
| if-eqz | instance-of |
| move-object | if-eqz |
| check-cast | check-cast |
| invoke-virtual | goto |
| move-result | new-instance |
| 5.  08 54 08 54 08 | 10.  08 54 6e 0c 08 |
| move-object/from16 | move-object/from16 |
| iget-object | iget-object |
| move-object/from16 | invoke-virtual |
| iget-object | move-result-object |
| move-object/from16 | move-object/from16 |

*Table 8. Top 10 5-gram opcode from malware detection.*

# BIOGRAPHICAL NOTES

**BooJoong Kang** is currently a Research Fellow at the Centre for Secure Information Technologies (CSIT) in Queen's University Belfast, United Kingdom. He received the B.S., M.S., and Ph.D. degrees in electronics and computer engineering from Hanyang University, Korea, in 2007, 2009, and 2013 respectively. He His research interests include malware detection/analysis, threat analysis, intrusion detection/prevention, cyber-physical resilience measures, and cloud security.

**Suleiman Y. Yerima** is a Research Fellow with the Centre for Secure Information Technologies (CSIT), Queen's University Belfast, UK. He received his Ph.D. in Mobile Computing and Communications in 2009 from University of South Wales (formerly Glamorgan), U.K. He holds an MSc in Personal, Mobile and Satellite Communications from the University of Bradford, U.K and a B.Eng. (First Class) degree in Electrical and Computer Engineering from Federal University of Technology (FUT) Minna, Nigeria. He is also a Certified Information Systems Security Professional (CISSP) and a Certified Ethical Hacker (CEH). His current research interests are in mobile security, malware analysis and detection, and authentication.

**Sakir Sezer** received the Dipl. Ing. degree in electrical and electronic engineering from RWTH Aachen University, Germany, and the Ph.D. degree in 1999 from Queen's University Belfast, U.K. Prof. Sezer is currently the Head of Network Security Research at Queen's University Belfast. His research is leading major advances in the field of high-performance content processing and is currently commercialized by Titan IC Systems. He is also cofounder and CTO of Titan IC Systems and a member of several executive committees.

**Kieran McLaughlin** received his M.Eng in electrical and electronic engineering and Ph.D. from Queen's University Belfast, UK, in 2003 and 2006 respectively. He is a Lecturer at the Centre for Secure Information Technologies (CSIT), where he leads research in cyber security for smart grids, industrial control systems (ICS), and supervisory control and data acquisition (SCADA) networks. His research interests include threat analysis, intrusion detection/prevention, and cyber-physical resilience measures.

**Reference** to this paper should be made as follows: Kang, B., Yerima, Y. S., Sezer, S. & McLaughlin, K. (2016). N-gram opcode analysis for Android malware detection. International Journal on Cyber Situational Awareness, Vol. 1, No. 1, pp231-255